\documentstyle[12pt]{article}

\advance\textheight by \topskip
\begin{document}
\tolerance=100000
\thispagestyle{empty}
\setcounter{page}{0}

\newcommand{\be}{\begin{equation}}
\newcommand{\ee}{\end{equation}}
\newcommand{\br}{\begin{eqnarray}}
\newcommand{\er}{\end{eqnarray}}
\newcommand{\ba}{\begin{array}}
\newcommand{\ea}{\end{array}}
\newcommand{\bi}{\begin{itemize}}
\newcommand{\ei}{\end{itemize}}
\newcommand{\bn}{\begin{enumerate}}
\newcommand{\en}{\end{enumerate}}
\newcommand{\bc}{\begin{center}}
\newcommand{\ec}{\end{center}}
\newcommand{\ul}{\underline}
\newcommand{\ol}{\overline}
\newcommand{\eebbww}{$e^+e^-\rightarrow b\bar b W^+W^-$}
\newcommand{\ar}{\rightarrow}
\newcommand{\sm}{${\cal {SM}}$}
\newcommand{\mssm}{${\cal {MSSM}}$}
\newcommand{\Dir}{\kern -6.4pt\Big{/}}
\newcommand{\Dirin}{\kern -10.4pt\Big{/}\kern 4.4pt}
\newcommand{\DDir}{\kern -7.6pt\Big{/}}
\newcommand{\DGir}{\kern -6.0pt\Big{/}}

\def\Ord{\buildrel{\scriptscriptstyle <}\over{\scriptscriptstyle\sim}}
\def\OOrd{\buildrel{\scriptscriptstyle >}\over{\scriptscriptstyle\sim}}
\def\pl #1 #2 #3 {{\it Phys.~Lett.} {\bf#1} (#2) #3}
\def\np #1 #2 #3 {{\it Nucl.~Phys.} {\bf#1} (#2) #3}
\def\zp #1 #2 #3 {{\it Z.~Phys.} {\bf#1} (#2) #3}
\def\pr #1 #2 #3 {{\it Phys.~Rev.} {\bf#1} (#2) #3}
\def\prep #1 #2 #3 {{\it Phys.~Rep.} {\bf#1} (#2) #3}
\def\prl #1 #2 #3 {{\it Phys.~Rev.~Lett.} {\bf#1} (#2) #3}
\def\mpl #1 #2 #3 {{\it Mod.~Phys.~Lett.} {\bf#1} (#2) #3}
\def\rmp #1 #2 #3 {{\it Rev.~Mod.~Phys.} {\bf#1} (#2) #3}
\def\ijmp #1 #2 #3 {{\it Int.~J.~Mod.~Phys.} {\bf#1} (#2) #3}
\def\sjnp #1 #2 #3 {{\it Sov.~J.~Nucl.~Phys.} {\bf#1} (#2) #3}
\def\xx #1 #2 #3 {{\bf#1}, (#2) #3}
\def\preprint{{\it preprint}}

\begin{flushright}
{\large DFTT 60/94}\\ 
{\large DTP/94/104}\\ 
{\rm October 1994\hspace*{.5 truecm}}\\ 
\end{flushright}

\vspace*{\fill}

\begin{center}
{\Large \bf 
Contributions of below--threshold decays to ${\cal {MSSM}}$
Higgs branching ratios.\footnote{Work supported in part by Ministero 
dell' Universit\`a e della Ricerca Scientifica.\\[4. mm]
E-mail: Stefano.Moretti@durham.ac.uk; W.J.Stirling@durham.ac.uk.}}\\[2.cm]
{\large 
S.~Moretti$^{a,c}$ and W.~J.~Stirling$^{a,b}$}\\[0.5 cm]
{\it a) Department of Physics, University of Durham,}\\
{\it South Road, Durham DH1 3LE, United Kingdom.}\\[0.5cm]
{\it b) Department of Mathematical Sciences, University of Durham,}\\
{\it South Road, Durham DH1 3LE, United Kingdom.}\\[0.5cm]
{\it c) Dipartimento di Fisica Teorica, Universit\`a di Torino,}\\
{\it and I.N.F.N., Sezione di Torino,}\\
{\it Via Pietro Giuria 1, 10125 Torino, Italy.}\\[0.75cm]
\end{center}

\vspace*{\fill}

\begin{abstract}
{\normalsize\noindent We calculate all the 
experimentally relevant branching ratios of the Higgs bosons of the
Minimal Supersymmetric Standard Model, paying
particular attention to the contributions from below--threshold
decays. We show that in some cases these can significantly
change the pattern of branching ratios calculated without
taking off--shell effects into account.}
\end{abstract}

\vspace*{\fill}
\newpage

\section{Introduction}
It is likely that the first experimental evidence 
for supersymmetry will come from studies of the Higgs sector.
In supersymmetric extensions of the Standard Model (\sm)
the Higgs sector invariably has a rich structure,
which in the case of the Minimal Supersymmetric Standard Model 
(${\cal MSSM})$ is also highly constrained  (see for example
Ref.~\cite{guide}). 
Finding one or more of these Higgs bosons is one of the major
goals of present and future high--energy colliders.

Many years of phenomenological studies 
\cite{LHC,SSC,LepII,NLC,ee500,LC92,JLC}
have produced several important
lessons for Higgs searches. The first is that almost
all decay channels have large \sm\ backgrounds. It is therefore
important to have very precise calculations for the Higgs production
cross sections and  branching
ratios, so that small signals can be unambiguously identified.
A second lesson is that `below--threshold' decays can be very important.
By this we mean Higgs decays via intermediate states in which one
or more particles are off--mass--shell.
A simple example of this is provided by the \sm\ Higgs
decay $\phi\ar Z^0Z^0\ar 4l^\pm$. In the `on--shell' approach, one
would calculate the branching ratio for this as
$BR(\phi\ar Z^0 Z^0) \times BR(Z^0\ar l^+l^-)^2$, with
$BR(\phi\ar Z^0 Z^0) =0 $ for $M_\phi < 2 M_{Z^0}$. This, however,
is too naive. The below--threshold `off--shell' decay $\phi\ar Z^0(Z^0)^*\ar 4l^\pm$
is non--negligible, and in fact provides a very important Higgs
signature for $O(130\ {\rm GeV}) < M_\phi < 2 M_{Z^0}$.

Although below--threshold decays have been extensively studied (see below)
for the \sm, to date there has been no equivalent study of the \mssm.
In fact all \mssm\ Higgs phenomenological studies have so far used 
branching ratios based on on--shell calculations (except for the 
trivial cases of $H\ar W^\pm(W^\mp)^*, Z^0(Z^0)^*$ which are easily
 obtained from the
corresponding \sm\ results). In this paper we present
a complete calculation\footnote{For simplicity we ignore all direct decays
involving supersymmetric particles}
 of all relevant \mssm\ Higgs branching ratios,
paying particular attention to below--threshold contributions.
Of course it is impossible to cover {\it all} regions 
of the \mssm\ parameter space. Our aim is to identify
those situations where below--threshold decays are important, and those
where simple on--shell calculations are sufficient. We illustrate
our results by numerical calculations using `typical' parameter values. 
Our aim is to provide the necessary theoretical tools for precision
\mssm\ Higgs phenomenology at future colliders.

The paper is organized as follows. In the next section we give
a comprehensive review of \sm\ and \mssm\ branching ratios. In Section 3
we present numerical results for some typical \mssm\ scenarios,
highlighting the differences between off--shell and on--shell calculations.
In Section 4 we present our conclusions.

\section{\sm\ and \mssm\ Higgs Branching Ratios}

The branching ratios of the \mssm\ Higgs bosons have been 
studied in various papers. 
A good review of the early works on this subject can be found in
Ref.~\cite{guide}, where all the most relevant formulae
for on--shell decays have been summarized.
These include the \sm--like two--body decays\footnote{Throughout this study
we use $\Phi$ to denote the set of neutral \mssm\ Higgs bosons
$h$, $H$ and $A$} $\Phi\ar f\bar f$ 
(where $f$ represents a generic massive fermion)
\cite{SMHff}, $H\ar W^\pm W^\mp,Z^0Z^0$ \cite{unitaritySM1_SMHVV}
at tree--level, and the one--loop induced decays 
$\Phi\ar Z^0\gamma,\gamma\gamma$ \cite{SMHVph} 
and $gg$ \cite{SMHgg}, generalized to the \mssm\
\cite{MSSMHvarious,MSSMHVph}; and the \mssm--specific
decays $H^\pm\ar f\bar f'$,
$A\ar Z^0h$, $H^\pm\ar W^\pm h$, $H\ar hh,AA$, 
$\Phi\ar \tilde\chi^\pm_i\tilde\chi^\mp_j$
(chargino decays) and $\Phi\ar \tilde\chi^0_i\tilde\chi^0_j$
(neutralino decays) \cite{MSSMHvarious}. At the time of Ref.~\cite{guide}
 only the tree--level
Higgs mass relations had been computed, and 
the on--shell decays $h\ar AA$ and $H\ar H^\pm H^\mp,
Z^0A$ were kinematically forbidden. We now know that they are 
allowed at one--loop 
and the corresponding formulae can be easily obtained,
for example from Ref.~\cite{MSSMHvarious}.
Finally, the one--loop induced decays of the charged Higgses
$H^\pm\ar W^\pm Z^0,W^\pm\gamma$ have also been studied \cite{MSSMHWV}.

Recently, higher--order corrections to  most  of 
these processes have been computed.
For the \sm, one can find 
in the literature the QCD \cite{SMHqqQCDcorr} 
and EW \cite{SMHffEWcorr,SMHtautauVVEWcorr,SMHttEWcorr,SMHffEWEWcorr}
radiative corrections,
and their interplay \cite{SMHbbQCDEWcorr,SMHffQCDEWcorr} 
for the $f\bar f$--channels. The EW corrections to the $W^\pm W^\mp,
Z^0Z^0$ decay rates \cite{SMHtautauVVEWcorr,SMHWWEWcorr,MSSMneutralEWcorr}, 
and both the QCD \cite{SMHVphQCDcorr,SMHggQCDcorr}
and  EW  \cite{MSSMneutralEWcorr} corrections 
to the $Z^0\gamma,\gamma\gamma,gg$ decay rates 
are also now available. A detailed and updated review of the higher--order
corrections computed to date within the \sm\ can be found in 
Ref.~\cite{corrreview}.

As at tree--level,  the 
higher--order \sm\ corrections can  easily be  extended to 
the \mssm\ case. For the \mssm--specific processes, 
both the QCD \cite{MSSMHfgQCDcorr} and the EW \cite{MSSMHfgEWcorr} 
corrections to the fermionic decays $H^\pm\ar f\bar f'$
of the charged Higgs bosons, and the EW corrections \cite{MSSMneutralEWcorr}
to the decays involving neutral bosons (such as 
$H\ar hh$ and $A\ar Z^0h$) 
have been computed.

Within the \mssm, a systematic analysis of all the interesting
decays of both neutral and charged Higgs bosons, with reference to 
the search strategies at the LHC proton--proton collider and
taking certain of the above results into account, 
has been presented
in Ref.~\cite{KZproc}, using tree--level formulae for masses and couplings. 
This analysis has been updated in Ref.~\cite{KZ} to  include new results 
 for the one--loop corrections to  the latter 
\cite{corrMH0iMSSM,corrMHMSSM}. Similar studies have been reported in 
Refs.~\cite{0LHCSSC,0pmLHCSSC1,0pmLHCSSC2}, 
and extended to the  LEP~I and LEP~II $e^+e^-$ colliders
in Ref.~\cite{0pmLEPLHCSSC}. 
For a NLC $e^+e^-$ collider the corresponding analysis has been performed in 
Ref.~\cite{0pmNLC}.

Apart from the processes $H\ar W^{\pm*}W^{\mp*},Z^{0*}Z^{0*}$ 
\cite{SMHVVoff}, the results on decay widths and branching ratios reported
in Refs.~\cite{KZ,0LHCSSC,0pmLHCSSC1,0pmLHCSSC2,0pmLEPLHCSSC,0pmNLC}
have been obtained for the case of above--threshold decays only.
It is our aim in this paper to generalize these results to include also
below--threshold decays.   In particular, we study
the two--body decay channels
\be\label{decayin}
\Phi\ar s\bar s,c\bar c, b\bar b,t^*\bar t^*,\mu^+\mu^-,\tau^+\tau^-,
\ee
\be
\Phi\ar W^{\pm*}W^{\mp*},Z^{0*}Z^{0*},Z^{0*}\gamma,\gamma\gamma,gg,
\ee
\be
H\ar h^{*}h^{*},\quad\quad 
H\ar A^{*}A^{*},\quad\quad 
A\ar Z^{0*}h^{*},
\ee
for the neutral Higgses $\Phi = H$, $h$ and $A$, and
\be
H^+\ar c\bar s,t^*\bar b,\mu^+\nu_\mu,\tau^+\nu_\tau,
\ee
\be\label{decayout}
H^+\ar W^{+*}h^{*},
\ee
for the charged Higgses $H^\pm$'s.
We do not consider (i)  the tree--level decays 
\be
H\ar Z^{0*}A^{*},\quad\quad 
H\ar H^{+*}H^{-*},\quad\quad
h\ar A^{*}A^{*},
\ee
since these are only possible in regions of
the \mssm\  ($M_{A},\tan\beta$) parameter space already excluded
by LEP I (for $m_t<200$ GeV) \cite{excldecays}, and (ii)
the one--loop induced decays 
\be
H^+\ar W^{+*}\gamma,\quad\quad\quad
H^+\ar W^{+*}Z^{0*},
\ee
since these have very small  branching ratios \cite{MSSMHWV}.

In calculating the decay rates for 
the processes (\ref{decayin})--(\ref{decayout}) we have adopted
the off--shell decay formulae presented in Ref.~\cite{SMHtt_HVVoff} for
$\Phi\ar W^* W^*,Z^{0*}Z^{0*}$ and in Refs.~\cite{SMHVphoff,MSSMHVphoff} 
for $\Phi\ar Z^{0*}\gamma$.
 For the other cases not involving decays to top
 quarks we have recalculated
the off--shell formulae at tree--level, tracing the
corresponding amplitudes squared and integrating analytically
over the appropriate off--shell phase space.
We then integrate
 numerically over all the possible virtualities
of the final--state particles for all the processes. This is in contrast
to the  procedure sometimes used  
of constraining one of the decay products to be on--shell
when the decaying Higgs boson mass 
exceeds its rest mass. Although this allows the width to 
be computed analytically, it can give misleading results near threshold.
Our procedure is only  to use such `integrated' analytic
 expressions well above threshold. 
We therefore proceed in the following way.
With $A$ the decaying particle and  $B^{(*)}$ and $C^{(*)}$ the 
decay products, which can be either on--shell ($B,C$) or off--shell ($B^*,C^*$),
we define as the decay width of the channel $A\ar B^{(*)}C^{(*)}$ 
the function
\begin{eqnarray}\label{g1}
\Gamma(A\ar B^{(*)}C^{(*)}) & = & 
\int\int_R    \
\frac{{\rm d}Q^2_B M_B\Gamma_B}
     {\pi[(Q^2_B-M^2_B)^2+(\alpha_B\Gamma_B)^2]} \times\nonumber \\ & &  
\frac{{\rm d}Q^2_C M_C\Gamma_C}
     {\pi[(Q^2_C-M^2_C)^2+(\alpha_C\Gamma_C)^2]} \times
\Gamma(A\ar B^*C^*),
\end{eqnarray}
where $M_{B(C)}$ and $\Gamma_{B(C)}$ are the mass and the width
of the particle $B(C)$ with four--momentum squared $Q^2_{B(C)}$,
respectively, 
and $\Gamma(A\ar B^*C^*)$ the off--mass--shell  decay width into  $B$ and $C$.
The two--dimensional integration region $R$
is defined by 
\be
Q_{B}\geq m_b,\quad Q_{C}\geq m_c,\quad
Q_B+Q_C \leq M_A ,
\ee
where $m_{b(c)}$ represents the sum of the rest masses
of the decay products of the particle $B(C)$.
The change of variables  \cite{Brown}
\begin{equation}
 Q^2_X-M^2_X=M_X\Gamma_X\tan\theta_X,\quad
\Longrightarrow
{\rm d}Q^2_X=
\frac{(Q^2_X-M^2_X)^2+\alpha^2_X\Gamma^2_X}{M_X\Gamma_X}\;
{\rm d}\theta_X,\quad X=B,C,
\end{equation}
then gives an integrand smoothly dependent on the integration variables.
For the numerical evaluation we use 
 VEGAS \cite{vegas}.

The expression given in (\ref{g1}) is a good approximation when
$M_X  \gg m_x $ ($X=B,C$) so that $\Gamma_X \propto M_X$. We 
therefore use it for decays to vector--vector, vector--scalar
and scalar--scalar pairs.
It is {\it not} however a good approximation for decays involving
the top quark, e.g. $ \Phi \ar t \bar t$ or $H^+ \ar t\bar b$. For such
decays we must perform an exact matrix element calculation for the
complete process, e.g.  $ \Phi \ar  b W^+ \bar b W^-$ or $H^+ \ar 
b W^+\bar b$. 

We are not interested here in studying in detail the dependence
of the processes  (\ref{decayin})--(\ref{decayout})
on all   possible higher--order corrections
to the widths and branching ratios.
Instead we wish to find out which below--threshold \mssm\ Higgs decays
are important.
We therefore include only the larger non--SUSY
QCD corrections (when known), and not the EW ones.
In any case, our procedure can readily be adapted
to include such corrections if necessary.
We have included the modifications due to QCD loops in the 
computations following Ref.~\cite{SMHqqQCDcorr} for the processes
$\Phi\ar q\bar q$, Ref.~\cite{MSSMHfgQCDcorr} for $H^\pm\ar 
q\bar q'$, Ref.~\cite{SMHVphQCDcorr} for $h\ar \gamma\gamma,Z^0\gamma$
and, finally, Ref.~\cite{SMHggQCDcorr} for $h\ar gg$.
In particular, since the QCD corrections to the top loops in 
$Z^0\gamma$, $\gamma\gamma$ and $gg$ decays,
as given in Refs.~\cite{SMHVphQCDcorr,SMHggQCDcorr}, are
valid for $M_{\Phi}<2m_t$, we have implemented these only
for the lightest \mssm\ neutral Higgs, i.e. $\Phi=h$.
We do not include any threshold effects due to the possible
formation of $t\bar t$--bound states in the one--loop induced processes,
when the Higgs mass $M_{\Phi}$ reaches a value $\approx2m_t$ 
\cite{threshold}.

There is a slight subtlety concerning the `running' of the fermion
masses. For the decays involving light quarks $\Phi \ar q \bar q$
($q=s,c,b$), the use of the running
quark mass $m_q^2(Q^2 = M_\Phi^2)$ takes into account large logarithmic
corrections at higher orders in QCD perturbation theory
\cite{SMHqqQCDcorr}. In principle, one could imagine using the
same procedure for  $\Phi \ar t \bar t$  in the limit $M_\Phi
\gg m_t$. In practice, however, we are interested in the case
of $M_\Phi/m_t  \sim {\cal O}(1)$. For the top--loop mediated
decay $\Phi \ar gg$, it is known that the higher--order QCD corrections
are minimized if the quark mass is defined at the pole of the
propagator, i.e. $m_t^2(Q^2 = m_t^2)$  \cite{running}. For the
{\it pseudoscalar} Higgs, the decay rate for $A \ar gg$ has a local
maximum at $M_A = 2 m_t$. To be  consistent, therefore, we must use
the {\it same} top mass ($m_t^2(Q^2=m_t^2)$) in the decay rate for $\Phi
\ar t \bar t$.

To avoid a proliferation of additional
parameters, we neglect small chargino and neutralino contributions
in the $\Phi\ar Z^0\gamma$, $\gamma\gamma$ decays.
To further  simplify the discussion, we assume a universal
soft supersymmetry--breaking mass \cite{corrMH0iMSSM,corrMHMSSM}
\be 
m_Q^2=m_U^2=m_D^2=m_{\tilde q}^2,
\ee
and negligible mixing in the $stop$ and $sbottom$ mass matrices,
\be
A_t=A_b=\mu=0.
\ee
Moreover, if we also neglect the $b$--quark mass in the formulae
of Refs.~\cite{corrMH0iMSSM,corrMHMSSM}, the one--loop corrected
masses of the ${\cal {MSSM}}$  neutral ${\cal {CP}}$--even 
Higgs bosons can be expressed in terms
of a single parameter $\epsilon$ \cite{0pmLEPLHCSSC}, given by
\begin{equation}\label{m2}
\epsilon = \frac{3e^{2}}{8\pi^{2} M^{2}_{W^\pm}\sin^2\theta_W}\;  m_{t}^{4}
\;  {\rm {ln}}\left( 1 +
\frac{{m}^{2}_{\tilde q}}{m_{t}^{2}} \right).
\end{equation}
Diagonalization of the mass--squared matrix leads to the expressions
\begin{eqnarray}\label{m1}
M^{2}_{h,H}& = & 
\frac{1}{2}[M^{2}_{A} + M_{Z^0}^{2} + \epsilon/\sin^{2}\beta] \nonumber \\
&   & \pm \left\{ [ (M^{2}_{A} - M^{2}_{Z^0})\cos2\beta + \epsilon/\sin^{2}\beta]^{2}
+(M^{2}_{A} + M^{2}_{Z^0})^{2}{\sin}^{2}2\beta \right\}^{1/2},
\end{eqnarray} 
while the mixing angle $\alpha$ in the ${\cal {CP}}$--even sector
is defined at one--loop by
\begin{equation}\label{m3}
\tan 2\alpha = \frac{(M_{A}^{2} + M_{Z^0}^{2}){\sin}2\beta}{(M_{A}^{2} - M_{Z^0}^{2})
{\cos2}\beta + \epsilon/{\sin}^{2}\beta},
\quad\quad -\frac{\pi}{2}<\alpha\leq0.
\end{equation}
For the ${\cal {MSSM}}$ charged Higgs masses we have maintained the 
tree--level relations
\begin{equation}
M_{H^\pm}^2=M_{A}^2+M_{W^\pm}^2,
\end{equation}
since one--loop corrections are quite small in comparison
to those  for the neutral Higgs bosons \cite{corrMHMSSM}.

For  the above choice of parameters 
we also use the leading one--loop corrected expressions
for the $Hhh$ and $HAA$ trilinear couplings, which enter
in the Higgs branching ratios  studied here, i.e.
\be
\lambda_{Hhh}=\lambda_{Hhh}^0+\Delta\lambda_{Hhh},
\ee
\be
\lambda_{HAA}=\lambda_{HAA}^0+\Delta\lambda_{HAA},
\ee
with the tree--level relations
\be
\lambda_{Hhh}^0=  
-\frac{igM_{Z^0}}{2\cos\theta_W}
[2\sin(\beta+\alpha)\sin2\alpha-\cos(\beta+\alpha)\cos2\alpha],
\ee
\be
\lambda_{HAA}^0=
\frac{igM_{Z^0}}{2\cos\theta_W}\cos2\beta\cos(\beta+\alpha),
\ee
and the one--loop corrections
\be
\Delta\lambda_{Hhh}=
-\frac{igM_{Z^0}}{2\cos\theta_W}
\frac{3g^2\cos^2\theta_W}{8\pi^2}
\frac{\cos^2\alpha\sin\alpha}{\sin^3\beta}
\frac{m_t^4}{m_{W^\pm}^4}
\left(3\log\frac{m_{\tilde q}^2+m_t^2}{m_t^2}
      -2\frac{m_{\tilde q}^2}{m_{\tilde q}^2+m_t^2}\right),
\ee
\be
\Delta\lambda_{HAA}=
-\frac{igM_{Z^0}}{2\cos\theta_W}
\frac{3g^2\cos^2\theta_W}{8\pi^2}
\frac{\sin\alpha\cos^2\beta}{\sin^3\beta}
\frac{m_t^4}{m_{W^\pm}^4}
\log\frac{m_{\tilde q}^2+m_t^2}{m_t^2},
\ee
as given in Ref.~\cite{KZ}.

In the numerical calculations presented in the following section
we adopt the following values for the electromagnetic coupling constant
and the weak mixing angle,
$\alpha_{em}= 1/128$ and  $\sin^2\theta_W=0.23$, respectively.
The strong coupling constant $\alpha_s$, which appears at leading
order in the $gg$ decay widths and also enters via the
QCD corrections,  has been evaluated
at two loops, with $\Lambda^{(4)}_{\overline{\rm MS}}=190$ MeV, and with
the number of active flavours $N_f$ (and the corresponding
$\Lambda^{(N_f)}_{\overline{\rm MS}}$, calculated according to the
prescription of Ref.~\cite{MARCIANO})
and scale $Q^2$ chosen
appropriately for the decay in question,
i.e. adopting the prescriptions of
Refs.~\cite{SMHqqQCDcorr,SMHVphQCDcorr,SMHggQCDcorr,MSSMHfgQCDcorr}.

For the gauge boson masses and widths we take
$M_{Z^0}=91.1$ GeV, $\Gamma_{Z^0}=2.5$ GeV, 
$M_{W^\pm}=M_{Z^0}\cos\theta_W$ and 
$\Gamma_{W^\pm}=2.2$ GeV, while for the fermion masses we take
$m_\mu=0.105$ GeV, $m_\tau=1.78$ GeV, $m_s=0.3$ GeV, $m_c = 1.4$ GeV,
 $m_b=4.25$ GeV
and $m_t=175$ GeV \cite{CDF}, with all widths equal to zero except
for $\Gamma_t$. We calculate this at tree--level
within the \mssm, using the expressions
(for the above values of $m_t$ and $m_b$) \cite{widthtopMSSM}
\begin{eqnarray} 
\frac{\Gamma(t\ar bH^+)}{\Gamma(t\ar bW^+)}&=&
\frac{\lambda(M_{H^\pm}^2,m_b^2,m_t^2)^{1/2}}
     {\lambda(M_{W^\pm}^2,m_b^2,m_t^2)^{1/2}}\times \\ \nonumber
&&\frac{(m_t^2+m_b^2-M_{H^\pm}^2)(m_t^2{\rm cotan}^2\beta+
m_b^2\tan^2\beta)+4m_t^2m_b^2}
       {M_{W^\pm}^2(m_t^2+m_b^2-2M_{W^\pm}^2)+(m_t^2-m_b^2)^2},
\end{eqnarray} 
and \cite{widthtopSM}
\begin{eqnarray} 
{\Gamma(t\ar bW^+)}&=&|V_{tb}|^2
\frac{G_Fm_t}{8\sqrt{2}\pi}\lambda(M_{W^\pm}^2,m_b^2,m_t^2)^{1/2}\times\\ \nonumber
&&     \left\{\left[1-\left(\frac{m_b}{m_t}\right)^2\right]^2
      +\left[1+\left(\frac{m_b}{m_t}\right)^2\right]
       \left(\frac{M_{W^\pm}}{m_t}\right)^2 
      -2\left(\frac{M_{W^\pm}}{m_t}\right)^4\right\},
\end{eqnarray} 
where $V_{tb}$ is the Cabibbo--Kobayashi--Maskawa mixing angle
 (here set equal to 1), $G_F={\sqrt{2}g^2}/{8M_{W^\pm}^2}$ is
the electroweak Fermi constant,
with $g=e/\sin\theta_W$ and $-e$ the electron charge, and $\lambda^{1/2}$
is the usual kinematic factor
\be
\lambda(M_a,M_b,M_c)^{1/2}=[M_a^2+M_b^2+M_c^2-2M_aM_b-2M_aM_c-2M_bM_c]^{1/2}.
\ee
The first generation of fermions and all neutrinos are taken to be
massless, i.e. $m_u=m_d=m_e=m_{\nu_e}=0$ and
$m_{\nu_\mu}=m_{\nu_\tau}=0$, with zero decay widths.
Finally, the universal supersymmetry--breaking squark mass is
in the numerical analysis to be $m_{\tilde q}=1$ TeV.

\section{Numerical results}

It is impractical to cover all possible regions of the
\mssm\ parameter space. We choose a representative set
of figures which (a) describes the decay channels
and mass ranges which are relevant to experiment, (b)
illustrates the importance of below--threshold decays,
and (c) allows a comparison with previous studies.
Thus Figs.~1 -- 4 show the branching ratios
for the $h$, $H$, $A$ and $H^\pm$ \mssm\ Higgs bosons
respectively, as a function of the mass of the decaying
particle. Following Ref.~\cite{KZ}, we choose two
representative values, $\tan\beta = 1.5, 30$, which give
significant differences in certain channels.
In comparing our figures with those of Ref.~\cite{KZ},
it should be remembered that we have chosen a top quark mass
($m_t = 175$ GeV) consistent with the recent CDF measurement.

Figure~1 shows the branching ratios of the lightest neutral
scalar \mssm\ Higgs boson $h$ as a function of $M_h$.
Note that for this value of $m_t$, the maximum values
of $M_h$ are  $98(129)$~GeV for $\tan\beta = 1.5(30)$, achieved
in the limit  $M_A \ar \infty$. As expected, the $b \bar b$ and $
\tau^+\tau^-$ channels dominate over essentially the complete mass
range. The $c \bar c$ and $t \bar t$ branching ratios are smaller
for larger $\tan\beta$, reflecting the dependence of the $hf \bar f$
couplings on the angles $(\alpha,\beta)$.

The new feature of Fig.~1 compared with the corresponding figure
(Fig.~10) of Ref.~\cite{KZ} is the appearance of the below--threshold
decays $h \ar W^+W^-,\ Z^0Z^0$, especially for smaller  $\tan\beta$ values.
In fact for $\tan\beta = 1.5$, the $W^+W^-$ branching ratio is larger
than that for $\gamma\gamma$ over a sizeable
portion of the $h$ mass range (when $M_h \OOrd 88$~GeV). 
Fortunately, because
of their negligible contribution to the total width compared to the
dominant modes, the two new channels do not depress 
the $\gamma\gamma$ branching ratio, which is very important for Higgs 
searches at the LHC proton--proton collider. 
This does  happen at $\tan\beta = 30$ 
(and also affects the  $b\bar b$ and $\tau^+\tau^-$ channels),
but  only in a very narrow mass window, and so is unlikely to be
phenomelogically relevant.
 Note that
the quasi--singular behaviour of the branching ratios near $M_h =
M_h^{\rm max}$ is caused by the mapping of a large range of $M_A$ masses
into a small range of $M_h$ masses, see Ref.~\cite{KZ} for a fuller
discussion.

In Fig.~2 we show the corresponding set of branching ratios for the heavier
of the two neutral scalar \mssm\ Higgs bosons $H$. This has a rather
complicated structure which is discussed in Ref.~\cite{KZ}.
In summary, for large $\tan\beta$ the only important decays
are to $b \bar b$ and $\tau^+\tau^-$, apart from a very small
mass region around $M_H = M_H^{\rm min}$ where other decay channels are
important. For smaller $\tan\beta$, either the $W^+W^-$, $hh$ or
$t \bar t$ channels dominate, depending on $M_H$. The
inclusion of  below--threshold decays has little effect here,
serving mainly to generate a smooth transition around
the $t \bar t$ threshold region.

Figure~3 shows the branching ratios for the pseudoscalar $A$ Higgs
boson. Here the difference between large and small $\tan\beta$ is
again very marked.
For the former, the $b \bar b$ and $\tau^+\tau^-$ decays
are completely dominant, and below--threshold decay effects
are negligible. For the latter, the $t \bar t$ decay
dominates above threshold and the $Z^0 h$ decay channel can also be important.
The point which we wish to stress is that the below--threshold
$t \bar t$ decay plays an important role. To see this more clearly,
Fig.~4  shows the corresponding set of branching ratios when
below--threshold decays are forbidden. (This is the analogue
of Fig.~12 in Ref.~\cite{KZ}.) In this case for $\tan\beta = 1.5$
there are three distinct
mass regions: $M_A < (M_{Z} + M_h)$, $(M_{Z} + M_h) < M_A < 2 m_t$
and $2 m_t < M_A$ where the dominant decay modes are $b \bar b$,
$Z^0 h$ and $ t \bar t$ respectively. However, the middle mass region
is strongly affected by the below--threshold $t \bar t$ decay.
In particular, for $300\ {\rm GeV} < M_A < 350\ {\rm GeV}$
the $b \bar b$ and $Z^0 h $ decay modes are suppressed far below
their `naive' values shown in Fig.~4. This clearly has important
phenomenological implications, especially since the leading
$A \ar t \bar t \ar b W^+ \bar b W^-$ channel is not readily  observable.

Another interesting effect is seen in the charged Higgs branching
ratios, Fig.~5. At large $\tan\beta$ the only relevant decays are
$tb$ and $\tau\nu$. Here the inclusion of the below--threshold
decay for the former simply gives a smoother transition between the two;
compare Fig.~14(b) of Ref.~\cite{KZ}. At smaller $\tan\beta$ there is
a window (assuming the CDF central value for $m_t$) for the $W h$
channel to be important. In fact this would be the dominant
channel over a sizeable range of $M_A$
in the absence of the below--threshold $tb$ decay. However, we
see that when the threshold behaviours are correctly taken into account
 the $tb$ decay rate  is almost always larger than that for the
 the $W h$ channel. Again, this has important implications for
charged Higgs searches at future colliders.

\section{Conclusions}

In summary, we have performed a comprehensive
 study of the phenomenologically
relevant decay rates of the \mssm\ Higgs bosons, paying particular
attention to below--threshold decay channels. We have discovered
 several cases
where these may be important. For $\tan\beta$ not too large, there 
are non-neglible branching ratios for the decay of the lightest neutral
Higgs boson $h$ into off-shell $W^+ W^-$ and $Z^0Z^0$. At the upper
end of the $M_h$ mass range these can reach ${\cal O}(10^{-3} - 10^{-2})$
and (in the case of  $W^+ W^-$)
can be larger than the important $\gamma\gamma$ branching ratio.

For the decays of heavier neutral bosons $H$ and $A$, the main impact 
is from below--threshold $t \bar t$ decays. Again for modest
values of $\tan\beta$ these can suppress the branching ratios of 
phenomenologically important channels such as $H\ar hh, W^+W^-, Z^0Z^0,
b \bar b$ and $A \ar Zh, b \bar b, \tau^+\tau^-$ below their
values calculated without taking the off-shell effects into account
(compare Figs.~3 and 4). In the same way, the below--threshold decay
$H^\pm \ar tb$ of the charged Higgs boson can suppress the branching ratio
of the important $H^\pm \ar W^\pm h$ channel.

As the physics studies for future high-energy colliders become 
more focused, it is important that all the ingredients
of `new physics' search strategies are calculated with as high a precision
as possible. We believe that our study of  \mssm\ Higgs boson decays
provides a significant improvement in this respect.

\section*{Acknowledgements}

\noindent Useful discussions with Ezio Maina are acknowledged.
 WJS is grateful to the UK PPARC for a  Senior Fellowship.

\section*{Figure captions}

\begin{itemize}
\item[{[1]}] Branching ratios for the lightest neutral scalar
\mssm\ Higgs boson $h$ as a function of $M_h$, for $\tan\beta = 1.5$
and 30. Other parameter values are given in the text.

\item[{[2]}] Branching ratios for the heavier of the two neutral scalar
\mssm\ Higgs bosons $H$ as a function of $M_H$, for $\tan\beta = 1.5$
and 30. Other parameter values are given in the text.

\item[{[3]}] Branching ratios for the pseudoscalar
\mssm\ Higgs bosons $A$ as a function of $M_A$, for $\tan\beta = 1.5$
and 30. Other parameter values are given in the text.

\item[{[4]}] As for Fig.~3, but with the below--threshold decays omitted.

\item[{[5]}] Branching ratios for the charged
\mssm\ Higgs bosons $H^\pm$ as a function of $M_{H^\pm}$, for $\tan\beta = 1.5$
and 30. Other parameter values are given in the text.

\end{itemize}

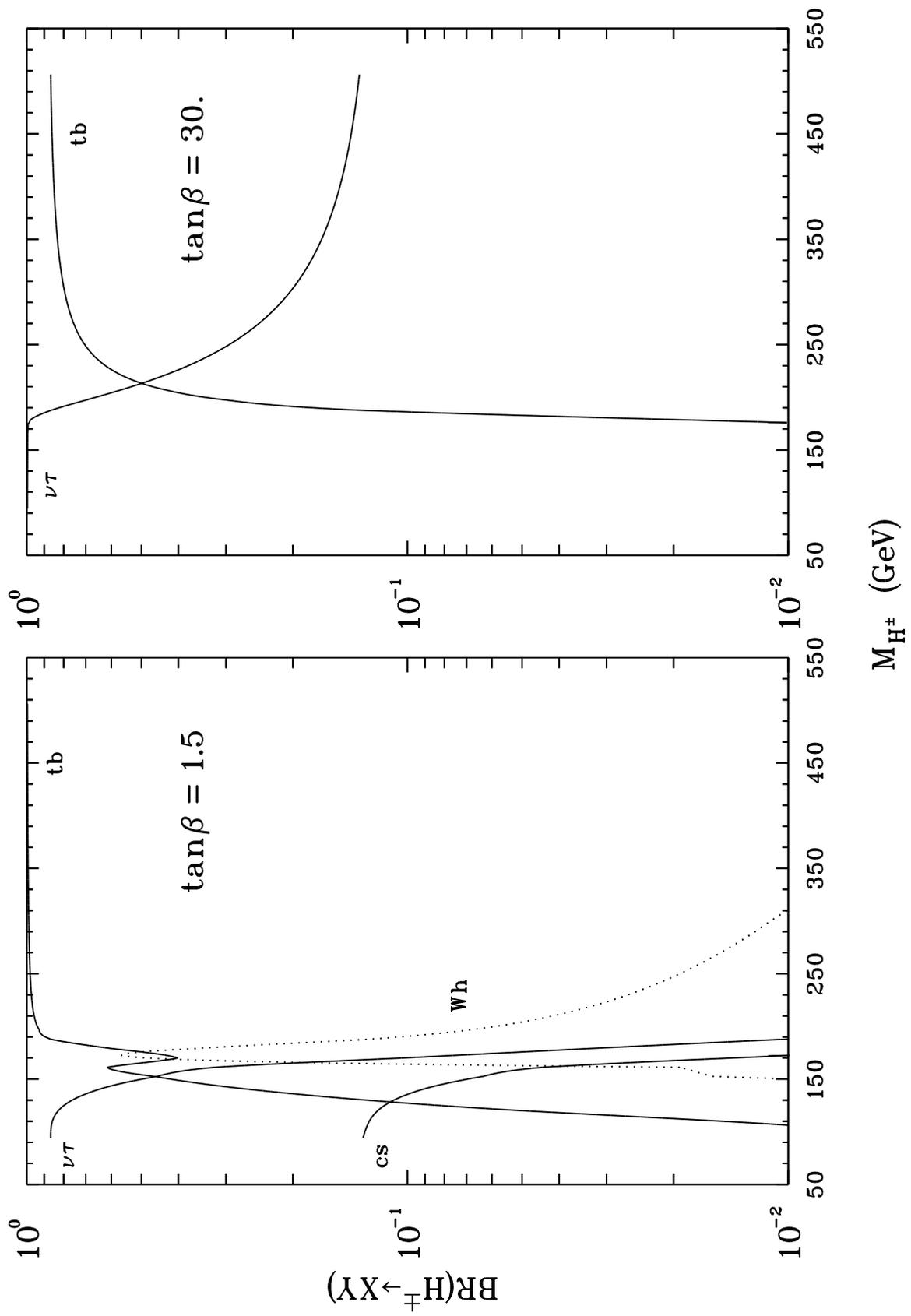
\end{document}